\documentclass[runningheads]{llncs}

\usepackage{graphicx}
\usepackage{booktabs}  %
\usepackage{subfig}
\usepackage[export]{adjustbox}
\usepackage{listings}

\usepackage{amsmath,amssymb}
\usepackage[]{microtype}
\usepackage{multirow}
\usepackage[colorlinks]{hyperref}
\usepackage{enumitem}
\usepackage{xspace}
\setlist[description]{style=nextline}

\hypersetup{
    colorlinks=true,
    linkcolor=blue,
    filecolor=magenta,
    urlcolor=blue,
    citecolor = blue,
}

\usepackage{caption}
\newcommand{\mldc}{\textsf{mldc}\xspace}
\newcommand{\mldclsids}{\textsf{mldc-lsids-phase}\xspace}

\newcommand{\chronobt}{\textsf{Maple\_LCM\_Dist\_ChronoBTv3}}

\usepackage{formatting/shortcuts}
    \title{Designing New Phase Selection Heuristics}
	\author{Arijit Shaw \and Kuldeep S. Meel}
	\institute{School of Computing, National University of Singapore}
\begin{document}

    \maketitle              %

\begin{abstract}
    CDCL-based SAT solvers have transformed the field of automated reasoning owing to their demonstrated efficiency at handling problems arising from diverse domains. The success of CDCL solvers is owed to the design of clever heuristics that enable the tight coupling of different components. One of the core components is phase selection, wherein the solver, during branching, decides the polarity of the branch to be explored for a given variable.
    Most of the state-of-the-art CDCL SAT solvers employ \emph{phase-saving} as a phase selection heuristic, which was proposed to address the potential inefficiencies arising from {\em far-backtracking}. In light of the emergence of chronological backtracking in CDCL solvers, we re-examine the efficiency of phase saving. Our empirical evaluation leads to a surprising conclusion:  The usage of saved phase and random selection of polarity for  decisions following a chronological backtracking leads to an indistinguishable runtime performance in terms of instances solved and PAR-2 score.

    We introduce Decaying Polarity Score (DPS) to capture the {\em trend} of the polarities attained by the variable, and upon observing lack of performance improvement due to DPS, we turn to a more sophisticated heuristic seeking to capture the activity of literals and the trend of polarities:   Literal State Independent Decaying Sum (LSIDS). We find the 2019 winning SAT solver, {\chronobt}, augmented with LSIDS solves 6 more instances while achieving a reduction of over 125 seconds in PAR-2 score, a significant improvement in the context of the SAT competition.

\end{abstract}

\section{Introduction}
\label{sec:introduction}
Given a Boolean formula $F$, the problem of Boolean Satisfiability (SAT) asks whether there exists an assignment $\sigma$ such that $\sigma$ satisfies $F$. SAT is a fundamental problem in computer science with wide-ranging applications including bioinformatics~\cite{lynce2006sat}, AI planning~\cite{kautz1992planning}, hardware and system verification~\cite{biere1999symbolic,clarke2004tool},  spectrum allocation, and the like. The seminal work of Cook~\cite{cook1971complexity} showed that SAT is NP-complete and the earliest algorithmic methods, mainly based on local search and the DPLL paradigm~\cite{davis1962machine}, suffered from scalability in practice. The arrival of Conflict Driven Clause Learning (CDCL) in the early '90s ~\cite{silva1996conflict} ushered in an era of sustained interest from theoreticians and practitioners leading to a medley of efficient heuristics that have allowed SAT solvers to scale to instances involving millions of variables~\cite{marques2009conflict}, a phenomenon often referred to as {\em SAT revolution}~\cite{audemard2009glucose,biere2013lingeling,een2003extensible,liang2017empirical,marques1999grasp,moskewicz2001chaff,nadel2018chronological,silva1996conflict}.

The progress in modern CDCL SAT solving over the past two decades owes to the design and the tight integration of the core components: {\em branching}~\cite{liang2016learning,silva2003grasp}, \emph{phase selection}~\cite{pipatsrisawat2007lightweight}, \emph{clause learning}~\cite{audemard2009predicting,luo2017effective}, \emph{restarts}~\cite{audemard2012refining,gomes1998boosting,huang2007effect,liang2018machine}, and \emph{memory management}~\cite{audemard2018glucose,oh2016improving}. The progress has often been driven by the improvement of the state of the art heuristics for the core components. The annual SAT competition~\cite{satcompetition} is witness to the pattern where development of the heuristics for one core component necessitates and encourages the design of new heuristics for other components to ensure a tight integration.

The past two years have witnessed the (re-)emergence of chronological backtracking, a regular feature of DPLL techniques, after almost a quarter-century since the introduction of non-chronological backtracking (NCB), thanks to Nadel and Ryvchin~\cite{nadel2018chronological}. The impact of chronological backtracking (CB) heuristics is evident from its quick adoption by the community, and the CB-based solver, \textsf{Maple\_LCM\_Dist \_ChronoBT}~\cite{ryvchin2018maple}, winning the SAT Competition in 2018 and and a subsequent version,  \textsf{Maple\_LCM\_Dist \_ChronoBTv3}  ,  the SAT Race 2019~\cite{satrace19} winner. The 2nd best solver at the SAT Race 2019, \textsf{CaDiCaL}, also implements chronological backtracking. The emergence of chronological backtracking necessitates re-evaluation of the heuristics for the other components of SAT solving.

We turn to one of the core heuristics whose origin traces to the efforts to address the inefficiency arising due to loss of information caused by non-chronological backtracking: the {\em  phase saving}~\cite{pipatsrisawat2007lightweight}  heuristic in the phase selection component. When the solver decides to branch on a variable $v$, the phase selection component seeks to identify the polarity of the branch to be explored by the solver. The idea of phase-saving traces back to the field of constraint satisfaction search~\cite{frost1994search} and SAT checkers~\cite{shtrichman2000tuning}, and was introduced in CDCL by Pipatsrisawat and Darwiche~\cite{pipatsrisawat2007lightweight} in 2007. For a given variable $v$, phase saving returns the polarity of $v$ corresponding to the last time $v$ was assigned (via decision or propagation). The origin of phase saving traces to the observation by Pipatsrisawat and Darwiche that for several problems, the solver may  forget a valid assignment to a subset of variables due to non-chronological backtracking and be forced to {\em re-discover} the earlier assignment. In this paper, we focus on the question: {\em is phase saving helpful for solvers that employ chronological backtracking? If not, can we design a new phase selection heuristic?}

The primary contribution of this work is a rigorous evaluation process to understand the efficacy of phase saving for all decisions following a chronological backtracking and subsequent design of improved phase selection heuristic. In particular,

\begin{enumerate}
    \item We observe that in the context of 2019's winning SAT solver \textsf{Maple\_LCM\_Dist \_ChronoBTv3} (referred to as {\mldc} henceforth)\footnote{acronyms in \textsf{sans serif} font denote solvers or solvers with some specific configurations. \textsf{mldc} is used as abbreviation for \textsf{\underline{M}aple\_{\underline{L}}CM\_{\underline{D}}ist\_{\underline{C}}hronoBTv3} }, phase saving heuristic for decisions following a chronological backtracking performs no better than the random heuristic which assigns positive or negative polarity randomly with probability 0.5.

    \item To address the inefficacy of phase saving for decisions following a chronological backtracking, we introduce a new metric, {\em decaying polarity score} (DPS), and DPS-based phase selection heuristic. DPS seeks to capture the {\em trend} of polarities assigned to variables with higher priority given to recent assignments. We observe that augmenting {\mldc} with DPS leads to almost the same performance as the default {\mldc}, which employs phase saving as the phase selection heuristic.

    \item To meet the dearth of performance gain by DPS, we introduce a sophisticated variant of DPS called \emph{Literal State Independent Decaying Sum}(LSIDS), which performs additive bumping and multiplicative decay. While LSIDS is inspired by VSIDS, there are crucial differences in computation of the corresponding activity of literals that contribute significantly to the performance. Based on empirical evaluation on SAT 2019 instances, {\mldc} augmented with LSIDS, called {\mldclsids} solves 6 more instances and achieves the PAR-2 score of 4475 in comparison to 4607 seconds achieved by the default {\mldc}.

    \item To determine the generality of performance improvement of {\mldclsids} over {\mldc}; we perform an extensive case study on the benchmarks arising from {\em preimage attack} on SHA-1 cryptographic hash function, a class of benchmarks that achieves significant interest from the security community.

\end{enumerate}

The rest of the paper is organized as follows. We discuss background about the core components of the modern SAT solvers in Section~\ref{sec:prelim}.  Section \ref{sec:phase-saving-eff} presents an empirical study to understand the efficacy of phase saving for decisions following a chronological backtracking. We then present DPS-based phase selection heuristic and the corresponding empirical study in Section \ref{sec:decay-pol}. Section~\ref{sec:lsids-phase-selection} presents the LSIDS-based phase selection heuristic. We finally conclude in Section~\ref{sec:conclusion}.

\section{Background}
\label{sec:prelim}

A literal is a propositional variable $v$ or its negation $\neg v$. A Boolean formula $F$ over the set of variables $V$ is in Conjunctive Normal Form (CNF) if $F$ is expressed as conjunction of clauses wherein each clause is a disjunction over a subset of literals. A truth assignment $\sigma: V \mapsto \{0,1\}$ maps every variable to $0$ (\textsc{False}) or $1$ (\textsc{True}).  An assignment $\sigma$ is called satisfying assignment or solution of $F$ if $F(\sigma) = 1$.  The problem of Boolean Satisfiability (SAT) seeks to ask whether there exists a satisfying assignment of $F$. Given $F$, a SAT solver is expected to return a satisfying assignment of $F$ if there exists one, or a proof of unsatisfiability~\cite{wetzler2013mechanical}.

\subsection{CDCL solver}\label{subsec:prelim:cdcl}

The principal driving force behind the so-called {\em SAT revolution} has been the advent of the Conflict Driven Clause Learning (CDCL) paradigm introduced by Marques-Silva and Sakallah~\cite{silva1996conflict}, which shares syntactic similarities with the DPLL paradigm~\cite{davis1962machine} but is known to be exponentially more powerful in theory. The power of CDCL over DPLL is not just restricted to theory, and its practical impact is evident from the observation that all the winning SAT solvers in the main track have been CDCL since the inception of SAT competition~\cite{satcompetition} .

On a high-level, a CDCL-based solver proceeds with an empty set of assignments and at every time step maintains a {\em partial assignment}. The solver iteratively assigns a subset of variables until the current partial assignment is determined not to satisfy the current formula, and the solver then backtracks while learning the reason for the unsatisfiability expressed as a conflict clause. The modern solvers perform frequent restarts wherein the partial assignment is set to empty, but information from the run so far is often stored in form of different statistical metrics.  We now provide a brief overview of the core components of a modern CDCL solver.

\begin{enumerate}
    \item \textbf{Decision.} The decision component selects a variable $v$, called the \emph{decision variable} from the set of unassigned variables and assigns a truth value, called the \emph{decision phase} to it. Accordingly,  a Decision heuristic is generally a combination of two different heuristics -- a \emph{branching heuristic} decides the decision variable and a \emph{phase selection heuristic} selects the decision phase.
    A \emph{decision level} is associated with each of the decision variables while it gets assigned. The count for decision level starts at $1$ and keeps on incrementing with every decision. %

    \item \textbf{Propagation.} %
    The propagation procedure computes the direct implication of the current partial assignment.
    For example, some clauses become \emph{unit} (all but one of the literals are \textsc{False}) with the decisions recently made by the solver. The remaining unassigned literal of that clause is asserted and added to the partial assignment by the propagation procedure. All variables that get assigned as a consequence of the variable $v$ get the same decision level as $v$.
    \item \textbf{Conflict Analysis.} Propagation may also reveal that the formula is not satisfiable with the current partial assignment. The situation is called a \emph{conflict}. The solver employs a {\em conflict analysis} subroutine to deduce the reason for unsatisfiability, expressed as a subset of the current partial assignment. Accordingly, the {\em conflict analysis} subroutine returns the negation of the literals from the subset as a clause $c$, called a \emph{learnt clause} or \emph{conflict clause} which is added to the list of the existing clauses. %
    The clauses in the given CNF formula  essentially imply the learnt clause.
    \item \textbf{Backtrack.} In addition to leading a learnt clause, the solver then seeks to undo a subset of the current partial assignment.  To this end, the conflict analysis subroutine computes the backtrack decision level $l$, and then the solver deletes assignment to all the variables with decision level greater than $l$.
    As the backtrack intimates removing assignment of last decision level only, backtracking for more than one level is also called \emph{non-chronological backtracking} or, \emph{backjumping}.
\end{enumerate}

The solver keeps on repeating the procedures as mentioned above until it finds a satisfying assignment or finds a conflict without any assumption.
The ability of modern CDCL SAT solvers to solve real-world problems with millions of variables depends on its highly sophisticated heuristics employed in different components of the solver.  Now we discuss some terms related to CDCL SAT solving that we use extensively in the paper.

\begin{itemize}

    \item \emph{Variable state independent decaying sum} (VSIDS) introduced in Chaff~\cite{moskewicz2001chaff} refers to a branching heuristic, where a score called $activity$ is maintained for every variable. The variable with the highest $activity$ is returned as the decision variable. Among different variations of VSIDS introduced later, the most effective is \emph{Exponential VSIDS} or, EVSIDS~\cite{biere2015evaluating,liang2015understanding} appeared in \textsf{MiniSat}~\cite{een2003extensible}. The EVSIDS score for variable $v$, $activity[v]$, gets incremented additively by a factor $f$ every time $v$ appears in a learnt clause. The factor $f$ itself also gets incremented multiplicatively after each conflict. A constant factor $g = 1/f$  periodically decrements the $activity$ of all the variables. The act of increment is called \emph{bump}, and the decrement is called \emph{decay}. The heuristic is called \emph{state independent} because the $activity$ of a variable is not dependent of the current state (e.g.,
    current assumptions) of the solver.
    \item \emph{Phase saving} \cite{pipatsrisawat2007lightweight} is a phase selection heuristic used by almost all solver modern solvers, with few exceptions such as the \textsf{CaDiCaL} solver in SAT Race 19~\cite{satrace19}. Every time the solver backtracks and erases the current truth assignment, phase saving stores the erased assignment. For any variable, only the last erased assignment is stored, and the assignment replaces the older stored assignment.
    Whenever the branching heuristic chooses a variable $v$ as the decision variable and asks phase saving for the decision phase, phase saving returns the saved assignment.
    \item \emph{Chronological backtracking.} When a non-chronological solver faces a \emph{conflict}, it \emph{backtracks} for multiple levels. %
    Nadel et al.~\cite{nadel2018chronological} suggested non-chronological backtracking (NCB) might not always be helpful, and advocated backtracking to the previous decision level. The modified heuristic is called \emph{chronological backtracking} (CB). We distinguish a decision based on whether the last backtrack was chronological or not. If the last backtrack is chronological, we say the solver is in \emph{CB-state}, otherwise the solver is in \emph{NCB-state}.

\end{itemize}

\subsection{Experimental Setup}
\label{subsec:experimental-setup}

In this work, our methodology for the design of heuristics has focused on the implementation of heuristics on a base solver and conduction of an experimental evaluation on a high-performance cluster for SAT 2019 benchmarks. %
We now describe our experimental setup in detail. All the empirical evaluations in this paper used this setup, unless mentioned otherwise.
\begin{enumerate}
    \item \textbf{Base Solver : } We implemented the proposed heuristics on top of the solver {\chronobt} {(\mldc)}, which is the winning solver for SAT Race 2019.  {\chronobt}  is an modification of \textsf{Maple\_LCM\_Dist\_ChronoBT  } (2018), which implements  chronological backtracking on top of \textsf{Maple\_LCM\_Dist} (2017).  \textsf{Maple\_LCM\_Dist}, in turn, evolved from \textsf{MiniSat} (2006) through \textsf{Glucose} (2009) and \textsf{MapleSAT} (2016). The years in parenthesis represent the year when the corresponding solver was published.

    \item \textbf{Code Blocks}: The writing style of this paper is heavily influenced from the presentation of \textsf{MiniSat} by E{\'e}n and S{\"o}rensson~\cite{een2003extensible}. Following E{\'e}n and S{\"o}rensson, we seek to present implementation details as code blocks that are intuitive yet detailed enough  to allow the reader to implement our heuristics in their own solver. Furthermore, we seek to present not only the final heuristic that performed the best, but we also attempt to present closely related alternatives and understand their performance.

    \item \textbf{Benchmarks : } Our benchmark suite consisted of the entire suite, totaling 400 instances, from SAT Race '19.
    \item \textbf{Experiments : } We conducted all our experiments on a high-performance computer cluster, with each node consists of E5-2690 v3 CPU with 24 cores and 96GB of RAM. We used 24 cores per node with memory limit set to 4GB per core, and all individual instances for each solver were executed on a single core. Following the timeout used in SAT competitions, we put a timeout of 5000 seconds for all experiments, if not otherwise mentioned. In contrast to SAT competition, the significant difference in specifications of the system lies in the size of RAM: our setup allows 4 GB of RAM in comparison to 128 GB of RAM allowed in SAT race '19.

    We computed the number of SAT and UNSAT instances the solver can solve with each of the heuristics. We also calculated the PAR-2 score. The PAR-2 score, an acronym for  penalized average runtime, used in SAT competitions as a parameter to decide winners, assigns a runtime of two times the time limit (instead of a ``not solved'' status) for each benchmark not solved by the solver.\footnote{All experimental data are available at \url{https://doi.org/10.5281/zenodo.3817476}.}
\end{enumerate}

\section{Motivation}
\label{sec:phase-saving-eff}
The impressive scalability of CDCL SAT solvers owes to the tight coupling among different components of the SAT solvers wherein the design of heuristic is influenced by its impact on other components. Consequently, the introduction of a new heuristic for one particular component requires one to analyze the efficacy of the existing heuristics in other components. To this end, we seek to examine the efficacy of phase saving in the context of recently introduced heuristic, Chronological Backtracking (CB). As mentioned in Section~\ref{sec:introduction}, the leading SAT solvers have incorporated CB and therefore, we seek to revisit the efficacy of other heuristics in light of CB. As a first step, we focus on the evaluation of phase selection heuristic.

\label{subsec:motiv-chronobt}

Phase saving was introduced to tackle the loss of precious work due to \emph{far-backtracking}~\cite{pipatsrisawat2007lightweight}.
Interestingly, CB was introduced as an alternative to {\em far-bactracking}, i.e., when the conflict analysis recommends that the solver should backtrack to a level $\hat{l}$ such that $|l - \hat{l}|$ is greater than a chosen threshold (say, $thresh$), CB instead leads the solver to backtrack to the previous level. It is worth noting that if the conflict analysis returns $\hat{l}$ such that $l - \hat{l} < thresh$, then the solver does backtrack to $\hat{l}$. Returning to CB, since the solver in CB-state does not perform far-backtracking, it is not clear if phase saving in CB-state is advantageous. To analyze empirically, we conducted preliminary experiments with {\mldc}, varying the phase-selection heuristics while the solver is performing CB. We fix the phase selection to phase saving whenever the solver performs NCB and vary the different heuristics while the solver performs CB:
\begin{enumerate}
    \item \emph{Phase-saving}  : Choose the saved phase as polarity, default in {\mldc}.

    \item \emph{Opposite of saved phase} : Choose the negation of the saved phase for the variable as polarity.
    \item \emph{Always false}: The phase is always set to \textsc{False}, a strategy that was originally employed in MiniSat 1.0.

    \item \emph{Random}           : Randomly choose between \textsc{False} and \textsc{True}.

\end{enumerate}

Our choice of {\em Random} among the four heuristics was driven by our perception that a phase selection strategy should be expected to perform better than {\em Random}. Furthermore, to put the empirical results in a broader context, we also experimented with the random strategy for both NCB and CB.  The performance of different configurations is presented in  \autoref{tab:chrono-polarity}, which shows a comparison in terms of the number of SAT, UNSAT instances solved, and   PAR-2 score.

\begin{table}[htb]
	\setlength{\tabcolsep}{8pt}
	\centering
	\begin{tabular}{cccccc}
		\toprule
		\multicolumn{2}{c}{Phase selection heuristic used} &  &  &  &  \\ \cline{1-2}
		In NCB-state & In CB-state & SAT & UNSAT & Total & PAR-2 \\  \midrule
        {\em Random} & {\em Random} & 133 & 89 & 222 & 5040.59 \\
		\emph{Phase-saving} & \emph{Phase-saving} & 140 & 97 & 237 & 4607.61 \\
        \emph{Phase-saving} & {\em Random} & 139 & 100 & 239 & 4537.65 \\
        \emph{Phase-saving} & \emph{Always false} & 139 & 98 & 237 & 4597.06 \\
        \emph{Phase-saving} & \emph{Opp. of saved phase} & 137 & 98 & 235 & 4649.13 \\ \bottomrule
	\end{tabular}
	\caption{Performance of {\mldc} on 400 \textsf{SAT19} benchmarks while aided with different phase selection heuristics. SAT, UNSAT, and total columns indicate the number of SAT, UNSAT, and SAT+UNSAT instances solved by the solver when using the heuristic. A lower PAR2  score indicates a lower average runtime, therefore better performance of the solver. }
	\label{tab:chrono-polarity}
\end{table}

We first observe that the {\mldc} solves 237 instances and achieves a PAR-2 score of 4607 -- a statistic that will be the baseline throughout the rest of the paper. Next, we observe that usage of random both in CB-state and NCB-state leads to significant degradation of performance: 15 fewer instances solved with an increase of 440 seconds for PAR-2. Surprisingly, we observe that random phase selection in CB-state while employing phase saving in NCB-state performs as good as phase-saving for CB-state. Even more surprisingly, we do not notice a significant performance decrease even when using {\em Always false} or {\em Opposite of saved phase}. These results strongly indicate that phase saving is not efficient when the solver is in CB-state, and motivate the need for a better heuristic. In the rest of the paper, we undertake the task of the searching for a better phase selection heuristic.

\section{Decaying Polarity Score for Phase Selection }
\label{sec:decay-pol}
To address the ineffectiveness of phase saving in CB-state, we seek to design a new phase selection heuristic while the solver is in CB-state. As a first step, we view phase saving as remembering only the last assigned polarity and we intend to explore heuristic design based on the recent history of polarities assigned to the variable of interest. Informally, we would like to capture the weighted {\em trend} of the polarities assigned to the variable with higher weight to the recently assigned polarity. To this end, we maintain a score, represented as a floating-point number, for every variable and referred to as  \emph{decaying polarity score}  (DPS). Each time the solver backtracks to level $l$, the assignments of all the variables with decision level higher than $l$ are removed from the partial assignment. We update the respective \emph{decaying polarity score} of all these variables, whose assignment gets canceled, using the following formula:

\begin{equation}
	\label{eqn:decay-pol-score}
	dps[v] = pol(v) + dec \times dps[v]
\end{equation}
where,

\begin{itemize}
	\item[-]  $dps[v]$ represent the \emph{decaying polarity score} of the variable $v$.
	\item[-] $pol(v)$ is $+1$ if polarity was set to \textsc{True} at the last assignment of the variable, $-1$ otherwise.
	\item[-] The decay factor $dec$ is chosen from $(0,1)$. The greater the value of $dec$ is, the more preference we put on polarities selected in older conflicts.
\end{itemize}

Whenever the branching heuristic picks a variable $v$ to branch on, the DPS-based phase selection heuristic returns positive polarity if $dps[v]$ is positive; otherwise, negative polarity is returned. Note that for $dec \leq 0.5$, the DPS-based heuristic is equivalent to phase saving modulo the discrepancies arising due to differences  in floating-point arithmetic and real number arithmetic.

\begin{figure}[t]
    \centering
    \parbox{0.75\textwidth}{
\begin{lstlisting}[language=c++, firstline=110, lastline=115]
void Solver.cancelUntil(int bLevel)  (* \hfill \itshape $\triangleright$ bLevel : backtrack level *)
(* \itshape - for all elements which are getting cancelled *)
  for (int c = trail.size()-1; c >= trailLim[bLevel]; c--)
    Var x = var(trail[c])
    dps[x] *= dec
    dps[x] += (sign(trail[c]) ? 1.0 : -1.0)
\end{lstlisting}}
    \parbox{0.75\textwidth}{
\begin{lstlisting}[language=c++, firstline=120, lastline=126]
Lit pickBranchLit()
  (*\itshape next : variable returned by branching heuristic *)
  if (dps[next] > 0)
    lit = mkLit(next, true)
  else
    lit = mkLit(next, false)
  return lit
\end{lstlisting}}
    \caption{Updating and using of decaying polarity score in \textsf{MiniSat} like code.}
    \label{fig:dec-pol-code}
\end{figure}

\subsection{Implementation}

The implementation of DPS-based heuristic closely resembles the implementation of the \emph{phase saving} heuristic \cite{pipatsrisawat2007lightweight}. Here we maintain an array of floating-point numbers named $dps$. Each of the elements in the array corresponds to the decaying polarity score of a variable in the formula. Whenever the solver backtracks and \emph{erases} the assignment for a variable $v$, the $dps[v]$ gets updated using \autoref{eqn:decay-pol-score}. At any later point during solving, if the branching heuristic decides to branch on the variable $v$,
the decaying polarity score heuristic returns a phase based on the value of $dps[v]$. The methods for updating $dps$ and picking up a literal based on $dps$ is shown in \autoref{fig:dec-pol-code}.

\subsection{Experimental Results}
To test the efficiency of DPS-based phase selection heuristic, we augmented\footnote{\url{https://github.com/meelgroup/duriansat/tree/decay-pol}} our base solver, {\mldc}, with DPS-based phase selection heuristic during CB. We set the value of $dec = 0.5$ and $0.7$ to understand the behavior with respect to varying values of $dec$.  As discussed in \autoref{subsec:experimental-setup}, we conducted our empirical evaluation on SAT-19 benchmarks.

\begin{table}[!t]
    \setlength{\tabcolsep}{12pt}
    \centering

    \begin{tabular}{ l c c c c }
        \toprule
        System & SAT & UNSAT & Total & PAR-2 \\ \midrule
        \textsf{mldc}  &  140 & 97 & 237 & 4607.61 \\
        \textsf{mldc-dec-phase-0.5} & 141 & 97 & 238 & 4589.66 \\
        \textsf{mldc-dec-phase-0.7} & 141 & 96 & 237 & 4604.66 \\ \bottomrule
    \end{tabular}
    \caption{Performance comparison of decaying polarity score with phase saving on \textsf{SAT19} instances. \textsf{mldc-dec-phase}\texttt{-<value>} represent the the solvers using decaying polarity score. The \texttt{<value>} represent the \emph{dec} used.}
    \label{tab:decay-score-result}
\end{table}

\paragraph{Solved Instances and PAR-2 score Comparison.}
\autoref{tab:decay-score-result} presents the comparison of the number of instances solved and PAR-2 score. We first note that the usage of DPS did not result in a statistically significant change in the performance of {\mldc}. Furthermore, the impact of $dec$ seems fairly limited as well. In light of the above remark of equivalence of phase saving and DPC with $dec \leq 0.5$, the floating-point computations introduce non-determinism due to lack of arbitrary precision. \autoref{tab:decay-score-result} indicates that the usage of DPS does not seem to enhance the aggregate performance of {\mldc}.
It is worth noting that there are significantly many instances where {\mldc} attains more than 20\% improvement over \textsf{mldc-dec-phase-0.7} and vice-versa. In 53 instances \textsf{mldc-dec-phase-0.7} had 20\% or more runtime improvement, while {\mldc} had 20\% or higher performance than \textsf{mldc-dec-phase-0.7} in 70 instances. The interesting behavior demonstrated by heuristic indicates, while  DPS-based phase selection heuristic fails to attain such an objective, it is possible to design heuristics that can accomplish performance improvement over phase saving. In the next section, we design a more sophisticated scheme that seeks to achieve the above goal.

\section{LSIDS : A VSIDS like heuristic for phase selection}
\label{sec:lsids-phase-selection}

We now shift to a more sophisticated heuristic that attempts to not only remember the trend of activity but also aims to capture the {\em activity} of the corresponding literal. To this end, we introduce a scoring scheme, called Literal State Independent Decay Sum (LSIDS), that performs additive bumping and multiplicative decay, {\`a} la VSIDS and EVISDS style. %
The primary contribution lies in the construction of policies regarding literal bumping. We maintain activity for every literal, and the activity is updated as follows:

\begin{enumerate}

	\item \textbf{Literal bumping} refers to incrementing $activity$ for a literal. With every \emph{bump}, the activity for a literal is incremented (additive) by  $inc*mult$, where $mult$ is a fixed constant while at every conflict, $inc$ gets multiplied by some factor $g > 1$. %
	\emph{Literal bumping} takes place in the following two different phases of the solver.
	\begin{description}
		\item[Reason-based bumping] When a clause $c$ is learnt, for all the literals $l_i$ appearing in $c$, the $activity$ for $l_i$ is bumped.  For example, if we learn the clause that consists of literals $v_5$, $\neg v_6$ and $v_3$, then we bump the $activity$ of literals $v_5$, $\neg v_6$  and $v_3 $.

		\item[Assignment-based bumping] While an assignment for a variable $v$ gets canceled during backtrack; if the assignment was \textsc{True}, then the solver bumps  $activity$ for $v$, otherwise the $activity$ for $\neg v$ is bumped.
	\end{description}

	\item \textbf{Literal decaying}  denotes the incident of multiplying the parameter $inc$ by a factor $> 1$ at every conflict. The multiplication of $inc$ implies the next bumps will be done by a higher $inc$. Therefore, the  older bumps to $activity$ will be relative smaller than the newer bumps. The name \emph{decaying} underscores the fact that the effect of increasing $inc$ is equivalent to decreasing (or, \emph{decay}-ing) the $activity$ of all literals.

	\item \textbf{Literal rescoring} : As the $activity$ gets incremented by a larger and larger factor every time, the value for $activity$ reaches the limit of a floating-point number at some time. At this point  the $activity$ of all the literals are scaled down.
\end{enumerate}

When the branching component returns a variable $v$, the LSIDS-based phase selection return positive if $activity[v] > activity[\neg v]$, and negative otherwise.
	One can view the proposed scheme as an attempt to capture both the participation of literals in learnt clause generation, in spirit similar to VSIDS, and storing the information about trend, {\`a} la phase saving/decay polarity score.

\begin{figure}[!t]
	\centering
	\parbox{0.5\textwidth}{
\begin{lstlisting}[language=c++, firstline=20, lastline=32]
void litBumpActivity(Lit l, double mult)
  activity[l] += inc * mult
  if (activity[l] > 1e100)
    litRescore()

void litDecayActivity()
  inc *= 1/decay

void litRescore()
  for (int i = 0; i < nVars(); i++)
    activity[i] *= 1e-100
    activity[(*$\neg$*)i] *= 1e-100
    inc *= 1e-100
\end{lstlisting}}
	\caption{Bump, Decay and Rescore procedures for LSIDS activity.}
	\label{fig:lsids-bump-decay}
\end{figure}

\subsection{Implementation Details}
\autoref{fig:lsids-bump-decay} shows the methods to bump and decay the LSIDS scores. \autoref{fig:places-to-call-bump} shows blocks of code from \textsf{MiniSat}, where the activity of literals is bumped. \autoref{fig:code_changes} showcases the subroutine to pick the branching literal based on LSIDS.  Of particular note is the setting of $mult$ to 2 for assignment-based bumping while setting $mult$ to 0.5 for Reason-based bumping. In order to maintain consistency with constants in EVSIDS, the constants in $litRescore$ are the same as that of EVSIDS employed in the context of branching in {\mldc}.

\begin{figure}[t]
	\centering
	\parbox{0.85\textwidth}{
\begin{lstlisting}[language=c++, firstline=40, lastline=57]
(* \scshape Bump literal scores for literals in learnt clause *)
void Solver.analyze(Constr confl)
 (* \itshape c : conflict clause *)
  litDecayActivity()
  for (int j =  0; j < c.size(); j++)
    Lit q = c[j];
    litBumpActivity((*$\neg$*)q, .5);

(* \scshape Bump literal scores when deleting assignment *)
void Solver.cancelUntil(int bLevel)  (* \hfill \itshape $\triangleright$ bLevel : backtrack level *)
 (* \itshape - for all elements which are getting cancelled *)
  for (int c = trail.size()-1; c >= trailLim[bLevel]; c--)
    Var x = var(trail[c])
    Lit l = mkLit(x, polarity[x])
    litBumpActivity((*$\neg$*)l, 2);


\end{lstlisting}}
	\caption{Sections in \textsf{MiniSat} like code, where LSIDS score is bumped and decayed.}
	\label{fig:places-to-call-bump}
\end{figure}

\begin{figure}[!t]
    \centering
    \parbox{0.8\textwidth}{
\begin{lstlisting}[language=c++, firstline=70, lastline=85]
Lit pickBranchLit()
  (* \itshape next : variable returned by branching heuristic *)
  (* \itshape CBT : denotes whether the last backtrack was chronological *)

  if (CBT)
      bool pol = pickLsidsBasedPhase(next)
      return mkLit(next, pol)
  else
      return mkLit(next, polarity[next])

bool pickLsidsBasedPhase(Var v)
  if ( activity[posL] > activity[negL] )
    return true
  else
    return false
\end{lstlisting}}
    \caption{Method to choose branching literal}
    \label{fig:code_changes}
\end{figure}

\subsection{Experimental Results}

To test the efficiency of LSIDS as a phase selection heuristic, we implemented\footnote{\url{https://github.com/meelgroup/duriansat/tree/lsids}} the heuristic on {\mldc}, replacing the existing phase saving heuristic.  We call the {\mldc} augmented with LSIDS phase selection heuristic as \mldclsids. %
Similar to the previous section; we tested the implementations on \textsf{SAT19} benchmarks using the setup mentioned in \autoref{subsec:experimental-setup}.

\begin{table}[!ht]
    \setlength{\tabcolsep}{12pt}
    \centering
    \begin{tabular}{lcccc}
        \toprule
        System & SAT & UNSAT & Total & PAR-2     \\ \midrule
        \textsf{mldc}    &  140 & 97 & 237 & 4607.61 \\
        \textsf{mldc-lsids-phase} & 147 &   96  &  243   & 4475.22 \\ \bottomrule

    \end{tabular}
    \caption{Performance comparison of LSIDS based phase selection with phase saving on 400 \textsf{SAT19} instances. }
    \label{tab:lsids-phase-sat}
\end{table}

\paragraph{Solved Instances and PAR-2 Score Comparison}
\begin{figure}[htb]
	\centering
    	{{\includegraphics[width=0.5\textwidth]{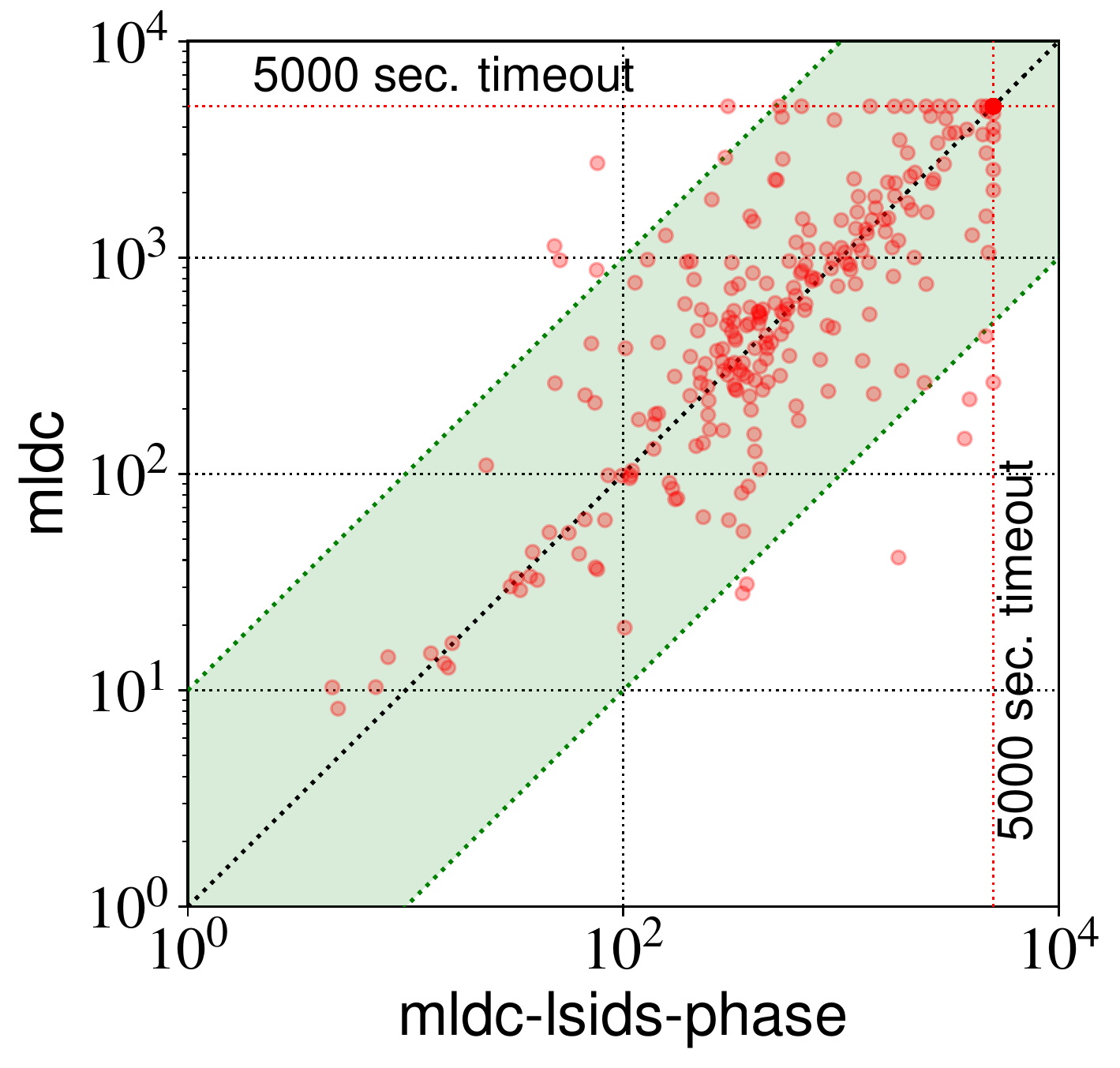} }}%
	\caption{  Performance comparison of {\mldclsids} vis-a-vis {\mldc}}
    \label{fig:lsids-plot-scatter}
\end{figure}
\begin{figure}[ht]
	\centering
	{\includegraphics[width=0.75\textwidth]{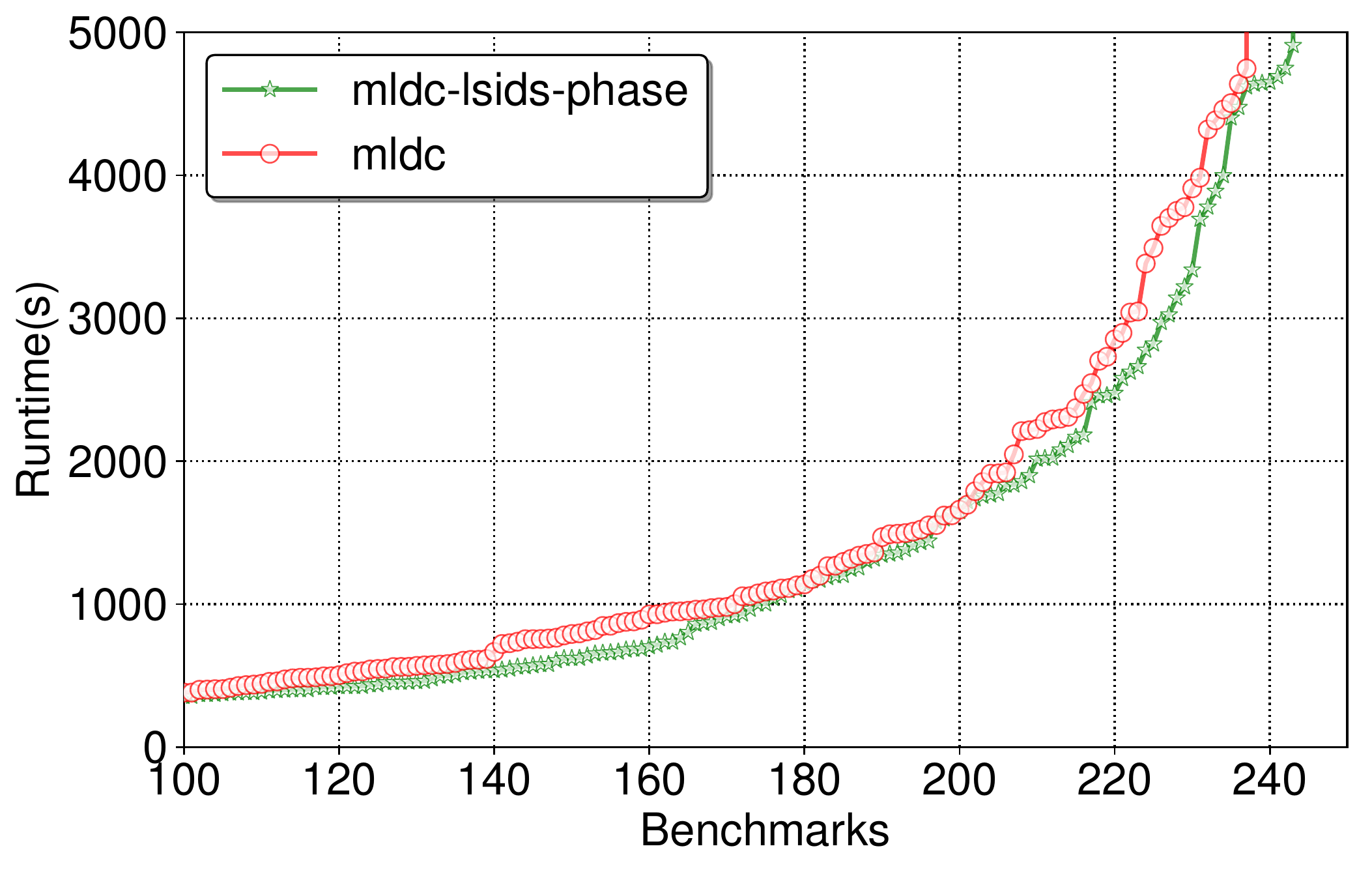} }%
	\caption{Each of the curve corresponds to the performance of a solver, by means of number of instances solved within a specific time.  At a specific runtime $t$, a curve to further right denotes the solver has solved more instances by time $t$.}.%
	\label{fig:lsids-plot-cactus}%
\end{figure}
 \autoref{tab:lsids-phase-sat} compares numbers of instances solved by the solver {\mldc} and {\mldclsids}. First, observe that {\mldclsids} solves 243 instances in comparison to 237 instances solved by {\mldc}, which amounts to the improvement of 6 in the number of solved instances. On a closer inspection, we discovered that {\mldc} performs CB for at least 1\% of backtracks only in 103 instances out of 400 instances. Since {\mldclsids} is identical to {\mldc} for the cases when the solver does not perform chronological backtracking, the improvement of 6 instances is out of the set of roughly 100 instances. It perhaps fits the often quoted paraphrase by Audemard and Simon~\cite{audemard2012refining}: {\em solving 10 or more instances on a fixed set (of size nearly 400) of instances from a competition by using a new technique, generally shows a critical feature}. In this context, we would like to believe that the ability of LSIDS-based phase selection to achieve improvement of 6 instances out of roughly 100 instances qualifies LSIDS-base phase saving to warrant serious attention by the community.

\autoref{tab:lsids-phase-sat} also exhibits enhancement in PAR-2 score due to LSIDS-based phase selection. In particular, we observe {\mldclsids} achieved reduction $2.87\%$ in PAR-2 score over {\mldc},  which is significant as per SAT competitions standards. In particular, the difference among the winning solver and runner-up solver for the main track in 2019 and 2018 was $1.27\%$ and $0.81\%$, respectively. In \autoref{fig:lsids-plot-scatter}, we show the scatter plot comparing instance-wise runtime performance of {\mldc} vis-a-vis {\mldclsids}. While the plot shows that there are more instances for which {\mldclsids} achieves speedup over {\mldc} than vice-versa, the plot also highlights the existence of several instances that time out due to the usage of {\mldclsids} but could be solved by {\mldc}.

\paragraph{Solving Time Comparison.} \autoref{fig:lsids-plot-cactus} shows a cactus plot comparing performance of {\mldc} and {\mldclsids} on \textsf{SAT19} benchmarks. We present number of instances on $x$-axis and the time taken on $y$-axis. A point $(x,y)$ in the plot denotes that a solver took less than or equal to $y$ seconds to solve $y$ benchmarks out of the $400$ benchmarks in \textsf{SAT19}. The curves for {\mldc} and {\mldclsids}  indicate, for every given timeout, the number of instances solved by {\mldclsids} is greater than or equal to {\mldc}.

\paragraph{Percentage of usage and difference of selected phase.}
Among the instances solved by \mldclsids, percentage of decisions taken with LSIDS phase selections is on average $3.17\%$ over the entire data set. %
Among the decisions taken with LSIDS phase selection, the average fraction of decisions where the selected phase differs from that of phase saving is $4.67\%$; It is worth remarking that maximum achieved is $88\%$ while the minimum is $0\%$.  Therefore, there are benchmarks where LSIDS and phase selection are entirely the same while there are benchmarks where they agree for only $12\%$ of the cases. The numbers thereby demonstrate that the LSIDS-based phase selection can not be simply simulated by random or choosing phase opposite of phase selection.

\paragraph{Applicability of LSIDS in NCB-state.}
The performance improvements owing the usage of LSIDS during CB-state raise the question of whether LSIDS is beneficial in NCB-state as well. To this end, we augmented {\mldclsids} to employ LSIDS-based phase selection during both NCB-state as well as CB-state. Interestingly, the augmented \mldclsids solved 228 instances, nine less compared to \mldc, thereby providing evidence in support of our choice of usage of LSIDS during CB-state only.

\paragraph{Deciding the best combination of CB and NCB.} Nadel and Ryvchin~\cite{nadel2018chronological} inferred that SAT solvers benefit from an appropriate combination of CB and NCB rather than solely reliance on CB or NCB. To this end, they varied two parameters, $T$ and $C$ according to the following rules to heuristically decide the best combination.
\begin{itemize}
    \item If the difference between the current decision level and backtrack level returned by conflict analysis procedure is more than $T$, then perform CB.
    \item For the first $C$ conflicts, perform NCB. This rule supersedes the above one.
\end{itemize}
Following the process, we experimented with different sets of $T$ and $C$ to determine the best combination of $T$ and $C$ for \mldclsids.  For each configuration ($T$ and $C$), we computed the performance of \mldc too. The results are summerized in \autoref{tab:lsids-phase-vary-t-c}. It turns out that $T = 100, C = 4000 $ performs best in \mldclsids. Interestingly, for most of the configurations , \mldclsids performed better than \mldc.

\begin{table}[!t]
    \setlength{\tabcolsep}{3pt}
    \centering
    \begin{tabular}{ll|cccc|cccc}
        \hline
        &  & \multicolumn{4}{c}{T = 100} & \multicolumn{4}{c}{C = 4000} \\ \cline{3-10}
        &  & C = 2000 & 3000 & 4000 & 5000 & T = 25 &  90 & 150 &  200 \\ \hline
        \multirow{2}{*}{SAT} & {\mldc} & 137 & \textbf{141} & 140 & 137 & 139 & 137 & 134 & 138 \\
        & {\mldclsids} & \textbf{143} & 139 & \textbf{147} & \textbf{139} & \textbf{142} & \textbf{141} & \textbf{139} & \textbf{142} \\ \hline
        \multirow{2}{*}{UNSAT} & {\mldc} & 98 & 96 & 97 & 97 & 98 & \textbf{97} & 95 & 97 \\
        & {\mldclsids} & \textbf{99} & \textbf{101} & \textbf{96} & \textbf{100} & \textbf{99} & 96 & \textbf{99} & 97 \\ \hline
        \multirow{2}{*}{Total} & {\mldc} & 235 & 237 & 237 & 234 & 237 & 233 & 229 & 235 \\
        & {\mldclsids} & \textbf{242} & \textbf{240} & \textbf{243} & \textbf{239} & \textbf{241} & \textbf{238} & \textbf{238} & \textbf{239} \\ \hline
        \multirow{2}{*}{PAR-2} & {\mldc} & 4663     & 4588 & 4607  & 4674 & 4609 & 4706 & 4773 & 4641 \\
        & {\mldclsids} & \textbf{4506} & \textbf{4558} & \textbf{4475} & \textbf{4575} & \textbf{4555} & \textbf{4556} & \textbf{4622} & \textbf{4583} \\ \bottomrule
    \end{tabular}
    \caption{Performance comparison of LSIDS based phase selection with phase saving on 400 \textsf{SAT19} instances with different $T$ and $C$.}
    \label{tab:lsids-phase-vary-t-c}
\end{table}

\subsection{Case Study on Cryptographic Benchmarks}
Following SAT-community traditions, we have concentrated on SAT-19 benchmarks. But the complicated process of selection of benchmarks leads us to be cautious about confidence in runtime performance improvement achieved by LSIDS-based phase selection. Therefore, in a bid to further improve our confidence in the proposed heuristic, we performed a case study on benchmarks arising from security analysis of {\em SHA-1 cryptographic hash functions}, a class of benchmarks of special interest to our industrial collaborators and to the security community at large. For a message $\mathcal{M}$, a cryptographic hash function $F$ creates a hash $\mathcal{H} = F(\mathcal{M})$. In a preimage attack, given a hash $\mathcal{H}$ of a message $\mathcal{M}$, we are interested to compute the original message $\mathcal{M}$. In the benchmark set generated, we considered SHA-1 with $80$ rounds, $160$ bits for hash are fixed, and $k$ bits out of $512$ message bits are fixed, $485 < k < 500$. The solution to the preimage attack problem is to give the remaining $(512 - k)$ bits.  Therefore, the brute complexity of these problems will range from $O(2^{12})$ to $O(2^{27})$. The CNF encoding of these problems was created using the SAT instance generator for SHA-1~\cite{nossum2013instance}. Note that by design, each of the instances is satisfiable.  In our empirical evaluation, we focused on a suite comprising of 512 instances\footnote{Benchmarks are available at \url{https://doi.org/10.5281/zenodo.3817476}.  } and every experiment consisted of running a given solver with $3$ hours of timeout on a particular instance.  %

\autoref{tab:lsids-phase-sha1} presents the runtime performance comparison of {\mldc} vis-a-vis {\mldclsids} for our benchmark suite. First, we observe that {\mldclsids} solves 299 instances in comparison to 291 instances solved by {\mldc}, demonstrating an increase of $8$ instances due to LSIDS-based phase selection. Furthermore, we observe a decrease of 229 in PAR-2 score, corresponding to a relative improvement of $2.30\%$, which is in the same ballpark as the improvement in PAR-2 score observed in the context of SAT-19 instances.

\begin{table}[!t]
    \setlength{\tabcolsep}{12pt}
    \centering
    \begin{tabular}{lcc}
        \toprule
        System & Total solved & PAR-2 \\ \midrule
         \textsf{mldc} &	291 & 9939.91 \\
        \mldclsids &	299 & 9710.42  \\ \bottomrule
    \end{tabular}
    \caption{Performance comparison of LSIDS based phase selection with phase saving on 512 cryptographic instances. Name of systems are same as \autoref{tab:lsids-phase-sat}.}

\label{tab:lsids-phase-sha1}
\end{table}

\section{Conclusion}\label{sec:conclusion}

In this paper, we evaluated the efficacy of phase saving in the context of the recently emerged usage of chronological backtracking in CDCL solving. Upon observing indistinguishability in the performance of phase saving vis-a-vis random polarity selection, 
we propose a new score: Literal State Independent Decay Sum (LSIDS) that seeks to capture both the activity of a literal arising during clause learning and also the history of polarities assigned to the variable. We observed that incorporating LSIDS to {\chronobt} leads to $6$ more solved benchmarks while attaining a decrease of 132 seconds in PAR-2 score.
The design of a new phase selection heuristic due to the presence of CB  leads us to believe that the community needs to analyze the efficiency of heuristics for other components in the presence of chronological backtracking.

\paragraph*{Acknowledgment}
We are grateful to the anonymous reviewers for constructive comments that significantly improved the final version of the paper. We are grateful to Mate Soos for identifying a critical error in the early drafts of the paper. 
This work was supported in part by National Research Foundation Singapore under its NRF Fellowship Programme[NRF-NRFFAI1-2019-0004 ] and AI Singapore Programme [AISG-RP-2018-005],  and NUS ODPRT Grant [R-252-000-685-13].  The computational work for this article was performed on resources of the National Supercomputing Centre, Singapore~\cite{nscc}.

\end{document}